# The effect of adding magnetic oxide as grain boundary for HAMR

Bing Zhou[1,2], B. S. D. Ch. S. Varaprasad[1,3], Zhengkun Dai[1,3], David E. Laughlin[1,2,3] and Jian-Gang Zhu[1,2,3]

[1]Data Storage Systems Center, Carnegie Mellon University, Pittsburgh, PA 15213, USA

[2]Materials Science and Engineering Department, Carnegie Mellon University, Pittsburgh, PA 15213, USA

[3]Electrical and Computer Engineering Department, Carnegie Mellon University, Pittsburgh, PA 15213, USA

**Grain-to-grain Curie temperature ($T_c$) variation in the media reduces signal-to-noise ratio due to its contribution in transition jitter noise, especially when average grain size is pushing down to increase the area storage capacity. A thermally insulating magnetic grain boundary may suppress such grain-to-grain $T_c$ variation, especially at small grain size. Here we present an experimental study on the effect of adding thermally-insulating magnetic oxide, in particular $BaFe_xO_y$, as part of the grain boundary materials in granular FePt-C HAMR media. It is found that the $BaFe_xO_y$ is chemically inert to FePt and the chemical ordering of FePt-$BaFe_xO_y$-C media are similar to that of FePt-C meida. By tuning the volume fraction of $BaFe_xO_y$ and C, well-separated FePt grains (average grain size = 6.8 nm) surrounded by $BaFe_xO_y$ shell with perpendicular $H_c$ above 35 kOe can be obtained. Transmission electron microscopy study with chemical analysis shows that the magnetic oxide appears to be crystalline and surround the FePt grains with immediate full enclosure. Magnetic measurements indicate an effective increase of magnetic grain size at temperatures below FePt Curie temperature. Pulsed laser pump-probe measurement indicates a measurable reduction of Curie temperature variation for the FePt-$BaFe_xO_y$-C media with carefully comparison with the reference FePt-C media.**

FePt-C granular film structure has been considered as the most promising candidate for heat-assisted magnetic recording (HAMR) media due to large magnetocrystalline anisotropy of $L1_0$ FePt with $K_u > 7 \times 10^7$ erg/cc.[1,2] During a recording process of HAMR, a transition position is determined when the medium is cooled from the Curie temperature to below the freezing temperature where the anisotropy field becomes beyond the capability of local head field. Grain-to-grain of $T_c$ variation, therefore, will directly translate into transition position jitter and can

yield significant degradation signal-to-noise ratio (SNR), limiting recording area density capability.[3–5]

The origin of the grain-to-grain Curie temperature variation arises from the following fact. For a FePt grain, ferromagnetic exchange fields on surface spins are significantly smaller than that of the spins in the core. When grain size becomes small, below 8nm, the Curie temperature of a FePt grain start to decrease with reducing grain size due to increase of surface-to-volume ratio. The resulting grain size dependence of the grain Curie temperature gives rise to Curie temperature distribution for a finite grain size distribution in the media.[6,7] The grain size dependence of the Curie temperature becomes more significantly pronounced at smaller grain sizes although reducing grain size is key for increasing recording area density capability for HAMR technology.

One way to mitigate the Curie temperature dependence on grain size is to let FePt-L10 grains to have magnetic grain boundaries of Curie temperature higher than that of FePt. This requires the magnetic grain boundary material to be thermally insulating to prevent lateral heat dissipation for maintaining the sufficient lateral thermal gradient in the magnetic medium layer during recording processes.

In this paper, we propose thermal-insulating magnetic oxide as part of grain boundary material for reducing grain-to-grain $T_c$ variation. In particular, we report the experimental study of mixing $BaFe_xO_y$ with C as the new grain boundary material. The possibility of utilizing $BaFe_{12}O_{19}$ was explored in this study, since it was found that $BaFe_{12}O_{19}$ is a ferrimagnetic material with high Curie temperature (can be > 800 K)[10] and relatively low thermal conductivity 4 $Wm^{-1}K^{-1}$.[11]

Films were sputtered on thermally grown $SiO_2$ substrates using an AJA Orion-8 system with base pressure lower than $2\times10^{-9}$ torr. Two series of samples C1-C3 and T1-T2 with film stack Si | $SiO_2$ | Ta (5 nm) | Cr (100 nm) | MgO(9 nm) | magnetic media (M) (7.5 nm) were prepared. The MgO, FePt, $BaFe_{12}O_{19}$ targets with purity of at least 99.9% have been used in this study. All samples have same structure up to MgO layer. In the first series of samples, the media layer M is FePt + X vol.% C, where X = 30, 32 and 35, for C1, C2 and C3, respectively. In the second series of samples, the media layer M is FePt + X vol.% (C + $BaFe_xO_y$), where X = 40 and 42, for

T1 and T2, respectively. Ta and Cr layer were DC sputtered at 5 mTorr at room temperature and 200 °C, respectively, and subsequently annealed in vacuum at 650 °C for 40 mins. Cr buffer layer is used to get large MgO grains with strong (002) texture and can be served as heat sink. MgO is RF sputtered at 10 mTorr at room temperature. Sample is then preheated at 650 °C for 30 mins to stabilize the deposition temperature of media layer M. FePt alloy target and C target were subsequently DC sputtered with RF-sputtered $BaFe_{12}O_{19}$ target onto pre-deposited MgO at 5 mTorr at 650 °C. The crystalline texture and degree of chemical ordering were analyzed and determined from X-ray diffraction spectra using Cu Kα radiation. The microstructure and chemical mapping of various samples were studied by bright-field transmission electron microscopy (TEM) imaging, high-resolution TEM imaging (HR-TEM), scanning TEM high angle annular dark field imaging (STEM-HAADF) and STEM energy dispersive x-ray spectroscopy (STEM-EDS). The FePt grain size analysis of all samples were examined from multiple regions of each sample in bright-field TEM. The room temperature magnetic properties and the moment (M) vs. temperature (T) behavior of all samples were examined using the Quantum Design's Magnetic Property Measurement System (MPMS 3) with a maximum field of 7 T. In M vs. T measurement, samples were arranged perpendicularly to the magnetic field direction. All samples were saturated at 7 T prior to performing the measurement at elevated temperature from 300 K to 700 K at zero field. The $T_c$ distribution of all samples were measured by pump probe setup based on the thermal remanence after pulse laser heating. The experimental set-up of $\sigma T_c$ measurement is mentioned in the previous literature with several modifications.[12] Compared with the previous setup, the pulse width was reduced from ~12 ns to ~0.7 ns, which further eliminates the effect of dwell time at high temperature. The pump beam and probe beam were combined and focused by 10x objective lens, which further reduced the spot size. Each pulse energy was independently measured by pulse energy meter. The measured $\sigma T_c$ of each sample was averaged from the fittings of five measurements at different locations.

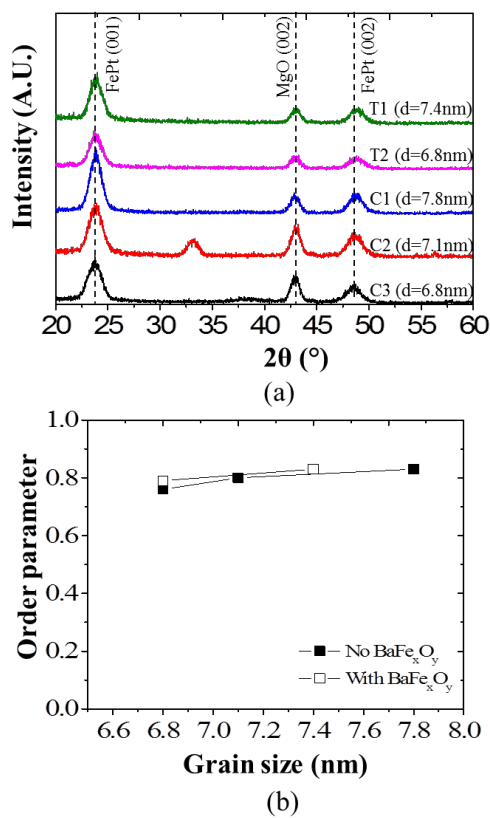

Fig. 1(a) XRD patterns of film stack Si | SiO$_2$ | Ta (5nm) | Cr (100nm) | MgO(9nm) | M (7.5nm) with M layer being FePt - X vol.% C (X=30, 32 and 35 for C1, C2 and C3) or FePt - X vol.% (C + BaFe$_x$O$_y$) (X=40 and 42 for T1 and T2); (b) Chemical order parameter (S) of C1-C3 and T1-T2 as a function of the average grain size.

Fig. 1 (a) shows the XRD patterns of C1-C3 and T1-T2, with MgO and FePt peaks indexed. Peak at 2θ=43° corresponds to MgO (002) and there are no other MgO peaks observed, suggesting that strong (002)-textured thin films were obtained. All samples show only FePt (001) and FePt (002) peaks indicating well-textured FePt grown on MgO with and without BaFe$_x$O$_y$. The large integrated intensity ratio between the super lattice peak FePt (001) to the fundamental peak FePt (002) in all samples demonstrate good chemical ordering in L1$_0$ FePt. The order parameter (S), shown in Fig. 1 (b), is calculated by considering the integrated intensity ratio of both peaks, the geometric features of XRD, the crystallographic texture of FePt films, and the film thickness.[13] S is plotted against the mean grain size of all samples, which was obtained from the in-plane bright field TEM images shown in the latter part. In the first series of samples, C1-C3 (represented by solid squares), S shows a decreasing trend, when the carbon volume percentage increases from 30 to 35. This is due to grain size reduction causing increasing surface to volume ratio of FePt grains as the carbon volume percentage increases. On the other hand, S of the second series of samples, T1 and T2 (represented by open squares), generally follows the same trend of that shown in C1-C3. The chemical ordering of FePt maintains with $BaFe_xO_y$ added in the media, suggesting that FePt is chemically inert to $BaFe_xO_y$.

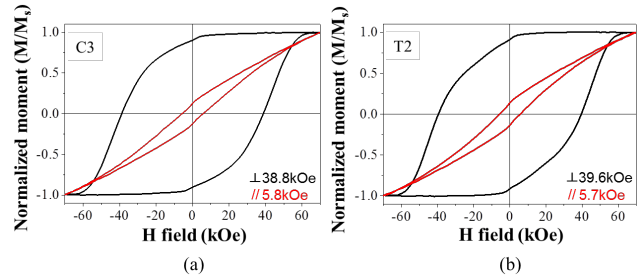

Fig. 2 Room temperature in-plane and perpendicular hysteresis loops for (a) C3 and (b) T2; in-plane and perpendicular $H_c$ as a function of average grain size for C1-C3 and T1-T2.

Fig. 2 (a)-(b) shows the typical room temperature perpendicular and in-plane M-H loops for samples with (T2) and without $BaFe_xO_y$ (C3), which have similar average grain size of 6.8nm. All samples show high perpendicular coercivity ranging from 38 kOe to 42 kOe, suggesting good $L1_0$ ordering of FePt films with and without $BaFe_xO_y$. The in-plane M-H loops of all samples show small opening with coercivity < 7 kOe and normalized remanence magnetization < 0.14, indicating the c-axis of FePt film are predominantly perpendicular to the film plane. Both perpendicular and in-plane magnetic behavior are consistent with the XRD results shown in Fig. 1. Note that the coercivity and squareness of the perpendicular M-H loop of sample C3 and T2 are similar, so it can be inferred that the $BaFe_xO_y$ in the media layer neither strongly exchange coupled the adjacent FePt grains nor destroy the microstructure.

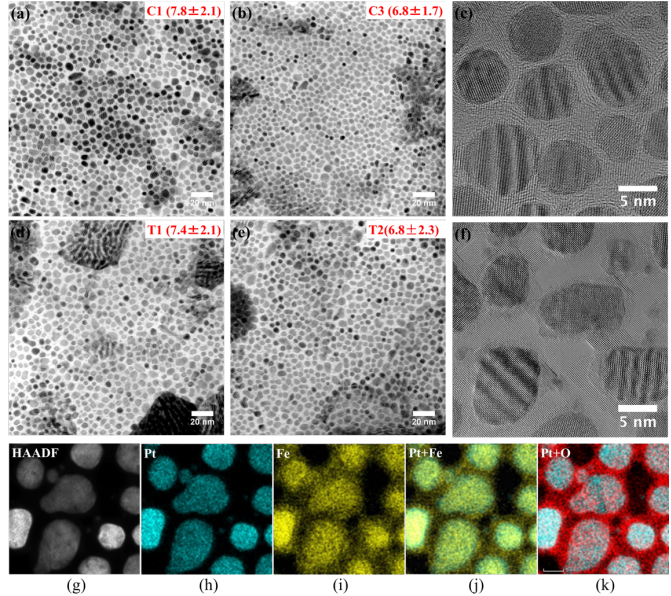

Fig. 3 The in-plane bright-field TEM micrographs for (a) C1, (b) C3, (d) T1 and (e) T2 with the insets being their corresponding average grain size and standard deviation; HR-TEM micrographs for (c) C1 (FePt - 30 vol.% C) and (f) T1 (FePt - 40 vol.% (C + BaFe$_x$O$_y$)); HAADF-STEM micrographs (g) and the EDS mapping of (h) Pt, (i) Fe, (j) Pt + Fe, and (k) Pt + O of T1 (FePt - 40 vol.% (C + BaFe$_x$O$_y$)).

Fig. 3 (a)-(b) and (d)-(e) shows the in-plane bright field TEM images of C1, C3, T1 and T2. In-plane bright field TEM image of C2 is not shown here. All samples show well-separated granular microstructure. The grain size analysis was performed on multiple regions from each sample to ensure the accuracy of the analysis. It is observed from our analysis that the mean grain size of FePt decreases gradually from 7.8 nm to 6.8 nm as the volume fraction of grain boundary material (C) increases from 30 vol. % to 35 vol. % in samples C1 to C3, whereas from T1 to T2, that of FePt decreases gradually from 7.4 nm to 6.8 nm as the volume fraction of grain boundary materials (C + BaFe$_x$O$_y$) increases from 40 vol. % to 42 vol. % of the total media layer. The in-plane microstructure of 7.5 nm FePt-30 vol.% BaFe$_x$O$_y$ was also examined (result not shown) and it shows an interconnected microstructure similar as other oxides such as SiO$_2$ and TiO$_2$ shown in previous literature.[14] Therefore, C is crucial to control the microstructure. Fig. 3 (c) and (f) shows the HR-TEM images for C1 and T1, respectively. It is observed that the grain boundary is amorphous for FePt-C media, whereas for FePt-BaFe$_x$O$_y$-C media, a crystalline shell is found around each FePt grain. To find where BaFe$_x$O$_y$ sits in the media, the elemental mapping of Fe, Pt and O was performed in the samples with BaFe$_x$O$_y$. The elemental mapping of Ba is not used in this study, since the atomic percentage of Ba is too low to be detected. Fig. 3 (g) to (k) shows the HAADF-STEM and EDS mappings of Fe and Pt of the FePt-BaFe$_x$O$_y$-C media,

T2. The chemical mapping of Pt shows as well-defined grains as FePt grains which suggest that the Pt is just in the FePt grains, which is consistent with the HAADF-STEM image. The chemical mapping of Fe shows Fe exists not only in FePt grains but also in the grain boundary, suggesting a Fe-rich grain boundary material is existing in the media. Fig. 3 (j) and (k) shows the overlapped EDS mappings of Fe-Pt and Pt-O. It clearly indicates that the crystalline shell surrounding FePt grains shown in HR-TEM images contains significant amount of Fe and O. Therefore, it can be concluded that the $BaFe_xO_y$ added in the media forms the crystalline shell surrounding FePt grains and shells are connected with each other. However, although $BaFe_xO_y$ shells formed around FePt grains are physically connecting together, the FePt grains are still magnetically decoupled from each other, since the room temperature perpendicular $H_c$, demonstrated in Fig. 2, are comparable for the cases with and without $BaFe_xO_y$ for similar mean grain size media. Based on the XRD patterns, the room temperature magnetic properties and the microstructure analysis, it can be concluded that the $BaFe_xO_y$ added in the media forms a crystalline shell surrounding FePt grains and it does not chemically interact with FePt nor strongly exchange couple the adjacent FePt grains.

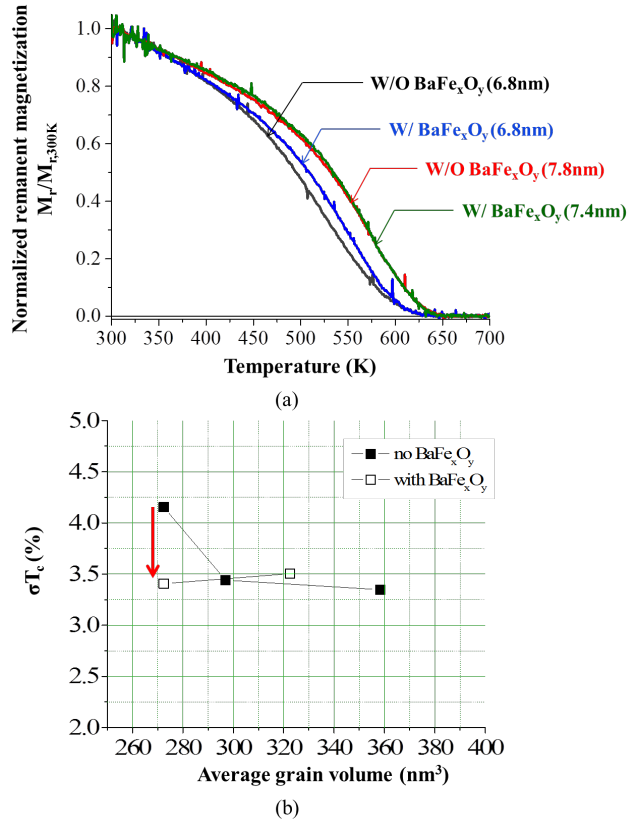

Fig. 4 (a) Experimental normalized $M_r$ vs. T curve for C1 (red), C3 (black), T1 (green) and T2 (blue); (b) Curie temperature distribution $\sigma T_C$ as a function of average grain volume for FePt-C and FePt-BaFe$_x$O$_y$-C media.

In the M vs. T measurement, the change of remanent magnetization ($M_r$) was measured when temperature increased from 300 K to 700 K. As the temperature increases, part of the FePt grains becomes super-paramagnetic and thus $M_r$ decreases. At certain temperature ($T_c^*$), $M_r$ drops to zero as all grains have become super-paramagnetic. Fig. 4 (a) shows the comparison of the normalized $M_r$ vs. T behavior between FePt-C media and FePt-BaFe$_x$O$_y$-C media. Note that $M_r$ at different temperature has been normalized with respect to its room temperature $M_r$ for each sample. Therefore, the moment contribution from the grain boundary has also been counted in. Red and black curves represent the $M_r$ vs. T behavior for FePt-C media with grain size of 7.8 nm (C1) and 6.8 nm (C3), whereas green and blue curves represent FePt-BaFe$_x$O$_y$-C with grain size of 7.4 nm (T1) and 6.8 nm (T2). In FePt-C media, the normalized $M_r$ vs. T curve of the sample with larger grain size (C1) shifts rightward toward higher temperature region compared with that of the sample with smaller grain size (C3), primarily due to finite size effect and surface effect. Comparing C3 and T2, it is evident that the addition of BaFe$_x$O$_y$ in the media shifts the normalized $M_r$ vs T curve towards higher temperature with same FePt grain size. The effect of adding $BaFe_xO_y$ in the media corresponds to a 5% increase of grain size in the FePt media with pure C grain boundary at 7.8 nm FePt grain case. This enhancement of $M_r$ at elevated temperature is due to the magnetic crystalline shell of $BaFe_xO_y$ formed around FePt grains, since the spins in the shell are weakly exchange coupled with the surface atomic spins of FePt grains to promote their coherence with those in the FePt core. As a result, the surface atomic spins of FePt at grain/magnetic grain boundary interface can maintain their spin coherence at a higher temperature compare with case of grain/non-magnetic grain boundary, such as FePt-C media. The exchange coupling between $BaFe_xO_y$ and FePt is weak enough so that the FePt grains are magnetically independent under magnetic field (shown in perpendicular hysteresis loop in Fig. 2) but strong enough to alleviate the FePt surface spin incoherence at grain/grain boundary interface.

Fig. 4 (b) shows the dependence of $\sigma T_C$ as a function of grain volume. The solid squares represent C1-C3 and open squares represents T1-T2. Each of the data point represents the best fit value from the thermo-remanence measurement of at least five different regions from each sample. The measured $\sigma T_C$ for FePt-C media C1-C3 increases significantly as the grain volume reduces and the trend is consistent with previous experimental work.[6,7] In comparison, $\sigma T_C$ of FePt-$BaFe_xO_y$-C media shows a 0.7% reduction in $\sigma T_C$ at smaller grain volume, which corresponds to around 20% reduction compared with FePt-C media. It can be inferred that the addition of $BaFe_xO_y$ in the media could have even larger effect at smaller grain size because of the hyperbolic increase in surface to volume ratio as the grain size reduces. The origin of this $\sigma T_C$ reduction at smaller grain volume is same as that causing enhancement of $M_r$ at elevated temperature, which has been discussed in the previous paragraph.

In summary, we have proposed to use thermally insulating magnetic oxide materials as part of grain boundary materials for granular FePt-L10 HAMR media for suppressing grain-to-grain $T_c$ variation. As a practical implementation of this concept, we have conducted a systematic experimental investigation with adding $BaFe_xO_y$ in granular FePt-C HAMR media. The study shows that co-sputtered $BaFe_xO_y$ stays together with carbon in the grain boundaries and the resulting microstructure are essentially the same as that of the FePt-C media. Moreover, the oxide in the grain boundaries forms crystalline shells, fully surrounding/enclosing FePt grains with lattice directly connected to FePt grains outer surfaces. The room temperature hysteresis measurements show that the perpendicular anisotropy field and coercivity are as high as that of the reference FePt-C samples, above 70 kOe and 35 kOe, respectively. XRD measurements show similar order parameters as well as similar degree of c-axis orientation distribution. The saturation remanence vs. temperature curve also indicates that the effective magnetic grain volume for $BaFe_xO_y$-C grain boundary media is greater than that of the reference C grain boundary media with similar FePt grain size below Curie temperature. Pulsed laser pump-probe measurements show that $\sigma T_C$ of FePt-$BaFe_xO_y$-C media with grain size of 6.8 nm shows around 20% reduction relative to that of FePt-C media with same grain size.

This work was supported in part by Western Digital, Seagate Technology, and the Data Storage Systems Center of Carnegie Mellon University. The authors acknowledge use of the Materials Characterization Facility at Carnegie Mellon University supported by grant MCF-677785.